# Hydrodynamic Time Scales and Temporal Structure of GRBs

Re'em Sari[1,2] and Tsvi Piran[1,3]

1. Racah Institute for Physics, The Hebrew University, Jerusalem, Israel 91904

2. Physics Department Nuclear Research Center- Negev, Beer-Sheva, Israel 84190

3. ITP, University of California, Santa Barbara, CA 93106, USA.

**Abstract**

We calculate the hydrodynamic time scales for a spherical ultra-relativistic shell that is decelerated by the ISM and discuss the possible relations between these time scales and the observed temporal structure in $\gamma$-ray bursts. We suggest that the bursts' duration is related to the deceleration time, the variability is related to the ISM inhomogeneities and precursors are related to internal shocks within the shell. Good agreement can be achieved for these quantities with reasonable, not fined tuned, astrophysical parameters. The difference between Newtonian and relativistic reverse shocks may lead to the observed bimodal distribution of bursts' durations.

## 1. Introduction

Gamma-ray bursts (GRBs) are most likely generated during deceleration of ultra-relativistic particles. A cosmological compact source that emits the energy re-quired for a GRB cannot generate the observed non thermal burst. Instead it will create an opaque fireball (Goodman, 1986; Paczyński, 1986; Piran, 1994). If even a small amount of baryonic matter is present then the ultimate result of this fireball will be a shell of ultra-relativistic particles (Shemi & Piran, 1990; Paczyński, 1990). The kinetic energy can be recovered as radiation only if these particles are decelerated by the ISM (Mészaros, & Rees, 1992) or by internal shocks (Rees & Mészaros 1994, Narayan, Paczyński & Piran, 1992). In retrospect, once this is realized, one can imagine GRB models in which the fireball is replaced by another (unknown) non-thermal acceleration mechanism but the radiation is still emitted due to slowing down of the ultra-relativistic particles. It is worthwhile, therefore, to explore the nature of the interaction between the ultra-relativistic particles and the ISM. We show that a careful analysis of this interaction changes some of the previous results and it may shed a new light on the expected temporal structure in GRBs.

We examine, first, in section 2 the planar shock problem. Spherical effects play, however, a crucial role in the realistic situation and we consider them in section 3. We discuss the observational implications to GRBs in section 4.

## 2. Planar Symmetry

Consider a slab of ultra-relativistic cold dense matter with a Lorentz factor $\gamma \gg 1$ that hits a stationary cold interstellar medium (ISM). Two shocks form: a reverse shock that propagates into the dense relativistic shell, reducing its speed and increasing its internal energy, and a forward shock that propagates into the Interstellar medium giving it relativistic velocities and internal energy. A contact discontinuity separates the shocked shell material and the shocked ISM material (Mészáros & Rees 1992; Katz, 1994).

There are four regions in this system: the ISM (1), the shocked ISM (2), the shocked shell material (3) and the unshocked shell material (4). The ISM is at rest relative to the observer. Velocities $\beta_i$, and their corresponding Lorentz factors $\gamma_i = (1 - \beta_i^2)^{-1/2}$, distances and time are measured relative to this frame. Thermodynamic quantities: $n_i$, $p_i$ and $e_i$ (particle number density, pressure and internal energy density) are measured in the fluids' rest frames. The ISM and the unshocked shell are cold and therefore: $e_1 = e_4 = 0$. The shocked material is extremely hot and therefore: $p_2 = e_2/3$ and $p_3 = e_3/3$.

For $\gamma \equiv \gamma_4 \gg 1$ the equations governing the shocks are (Blandford & McKee, 1976):

$$e_2/n_2 m_p c^2 = \gamma_2 - 1 \cong \gamma_2 \quad ; \quad n_2/n_1 = 4\gamma_2 + 3 \cong 4\gamma_2 , \tag{1}$$

$$e_3/n_3 m_p c^2 = \bar{\gamma}_3 - 1 \quad ; \quad n_3/n_4 = 4\bar{\gamma}_3 + 3 , \tag{2}$$

where $m_p$ is the protons rest mass. The approximations in equation 1 used only the fact that $\gamma_4 \gg 1$ and therefore $\gamma_2 \gg 1$. No assumption was made about $\bar{\gamma}_3$, the Lorentz factor of the motion of the shocked material in region 3 relative to the unshocked shell in region 4.

Equality of pressures and velocities along the contact discontinuity yields:

$$e_2 = e_3 \quad ; \quad \bar{\gamma}_3 \cong (\gamma_4/\gamma_2 + \gamma_2/\gamma_4)/2 \tag{3}$$

The solution for $\gamma_2$ depends only on two parameters $\gamma$ and $f \equiv n_4/n_1$. The energy, pressure and density also depend linearly on a third parameter, the external density $n_1$. A forth parameter, $\Delta$, the width (in the observer's frame) of the ultra-relativistic shell determines the time it takes the reverse shock to cross the shell, $t_\Delta$:

$$t_\Delta = \frac{\Delta}{c(\beta_4 - \beta_2)}\left(1 - \frac{\gamma n_4}{\gamma_3 n_3}\right) . \tag{4}$$

There are two simple limits of equations 1-4 in which the reverse shock is either Newtonian or ultra-relativistic (the forward shock is always ultra-relativistic if $\gamma \gg 1$ and $f > 1/\gamma^2$). If $\gamma^2 \gg f$ the reverse shock is ultra-relativistic ($\bar{\gamma}_3 \gg 1$) and:

$$\bar{\gamma}_3 = \frac{\gamma^{1/2}}{\sqrt{2}f^{1/4}} \quad ; \quad \gamma_2 = \gamma_3 = \frac{\gamma^{1/2}f^{1/4}}{\sqrt{2}} \tag{5}$$

In this case almost all of the initial kinetic energy is converted by the shocks into internal energy ($\gamma_3 \ll \gamma$). Therefore the process is over after a single passage of the reverse shock through the shell. The relevant time scale for energy extraction is the shell crossing time:

$$t_\Delta = \Delta\gamma\sqrt{f}/2c . \tag{6}$$

The internal energy densities in the shocked shell and in the shocked ISM is the same (see equation 3) and since both shocked regions have comparable width they release comparable amounts of energy. The ISM mass swept by the forward shock at the time that the reverse shock crosses the shell is $\sim f^{-1/2}$ of the shell's mass. This is larger than the simple estimate given by Mészaros & Rees (1992) of $\sim \gamma^{-1}$. At an earlier time when a mass of $\gamma^{-1}$ was swept, the reverse shock interacted only with a small fraction of the shell ($\sim \sqrt{f}/\gamma \ll 1$) and most of the energy was still the kinetic energy of the unshocked shell.

If $f \gg \gamma^2$ the reverse shock is Newtonian ($\bar{\gamma}_3 - 1 \ll 1$) and:

$$\bar{\gamma}_3 - 1 \cong 4\gamma^2 f^{-1}/7 \equiv 2\varepsilon \ll 1 \quad ; \quad \gamma_2 = \gamma_3 = \gamma(1 - \sqrt{\varepsilon}) . \tag{7}$$

The shock crosses the shell at:

$$t_\Delta = \sqrt{9/14}\Delta\gamma\sqrt{f}/c, \tag{8}$$

which is surprisingly similar (up to a constant) to the ultra-relativistic limit expression.

The reverse shock converts only a fraction $\gamma/\sqrt{f} \ll 1$ of the kinetic energy into internal energy. It is too weak to slow down the shell effectively and most of the initial energy is still kinetic energy when this shock reaches the inner edge of the shell. At this stage a rarefaction wave begins to propagate towards the contact discontinuity. This wave propagates at the speed of sound $\sqrt{4p_3/3n_3m_p}$ and it reaches the contact discontinuity at $t_r = (3\sqrt{7}/4)\Delta\gamma\sqrt{f}/c$, which is of the same order of magnitude as the shock crossing time $t_\Delta$. It is then reflected from the contact discontinuity and a second, weaker, shock wave forms. A quasi-steady state slowing down solution forms after a few crossings like this (Sari & Piran, 1995). Using momentum conservation, the total slowing down time can be estimated by $\sim \Delta\gamma n_4 m_p c/p_2 \sim \Delta f/c\gamma$. During this time the forward shock collects a fraction $\sim \gamma^{-1}$ of the shell's rest mass, which is the same as the original estimate of Mészáros & Rees (1992). In contrary to the relativistic case, there are two time scales now: the shock (or rarefaction) crossing time, $t_\Delta$, and the total slowing down time.

In the realistic situation the ISM density is probably inhomogeneous. Consider a density jump by a factor $f'$ over a distance $l_{ISM}$. The forward shock propagates into the ISM with a density $n_1$ as before and when it reaches the position where the ISM density is $n_1 f'$ a new shock wave is reflected. The solution of this problem requires the application of equations similar to equations 1-4. This shock is reflected again of the shell. Similar analyses shows that the reflections time is $\sim l_{ISM}/4c\sqrt{f'}$ and after these reflections the pressure and time scales are as if the ISM was homogeneous with a density $n_1 f'$ (Sari & Piran, 1995).

Finally, we mention the possibility of internal shocks inside the shell (Rees & Mészaros 1994; Narayan, Paczyński & Piran, 1992). These may form when faster material overtakes slower material. If the Lorentz factor varies by a factor of $\sim 2$ over a length scale $\delta R \leq \Delta$ then the time for these shock to from is $\sim \delta R \gamma^2 / c < \Delta \gamma^2 / c$. This time scale is shorter than the slowing-down time scale and therefore internal shocks appear before considerable deceleration in the Newtonian case. In the relativistic case considerable deceleration occurs before internal shocks unless $\delta R \ll \Delta$.

## 3. Spherical Considerations

The main difference between spherical and planar symmetries is that in a spherical system the density ratio $f \equiv n_4/n_1$ decreases with time. Initially $f/\gamma^2 \gg 1$ and the reverse shock is Newtonian. The energy conversion depends critically on the question whether this shock become relativistic before the kinetic energy is extracted from the shell. This depends on the ratio of two radii: $R_N$ where $f/\gamma^2 = 1$ and the reverse shock becomes relativistic and $R_\Delta$ where the reserve shock crosses the shell. Two other important radii are: $R_\gamma$ where the forward shock sweeps a mass $M/\gamma$ ($M$ is the shell's rest mass) and $R_s$ where the shell begins to spread if the initial Lorentz factor varies by order $\gamma$ (Mészaros, Laguna, & Rees, 1993; Piran, Shemi & Narayan, 1993; Piran, 1994). Note that $R_s$ is also an upper limit for the the location of internal shocks since $\delta R < \Delta$.

There are two intrinsic length scales in this problem. The first is $\Delta$, the width of the relativistic shell. The second is the Sedov length, $l \equiv (E/n_1 m_p c^2)^{1/3}$, which is familiar from SNR theory. The ISM rest mass within $l^3$ equals $E/c^2$. Typical GRB parameters yield, $l \approx 10^{18}$cm, which is similar to SNR value. We estimate the mass of the shell using, $M = E/\gamma c^2$, and obtain $R_\gamma = l/\gamma^{2/3}$. For a shell that propagates with a constant width $f \propto R^{-2}$ and we can express the other radii (omitting here and in the rest of the discussion factors of order of unity) in terms of $l$, $\Delta$ and $\gamma$: $R_N = l^{3/2}/\Delta^{1/2}\gamma^2$; $R_\Delta = l^{3/4}\Delta^{1/4}$ and $R_s = \Delta\gamma^2$. Conveniently, the four critical radii are related by one dimensionless quantity:

$$\xi \equiv (l/\Delta)^{1/2}\gamma^{-4/3} . \tag{9}$$

and

$$R_N/\xi = R_\gamma = \sqrt{\xi}R_\Delta = \xi^2 R_s . \tag{10}$$

Two possibilities exist:

1. $\xi > 1$ - the Newtonian case: $R_s < R_\Delta < R_\gamma < R_N$ and shock reaches the inner edge of the shell while it is still Newtonian. Most of the energy is extracted during a steady state deceleration phase described in the previous section. Since $R_s$ is smaller than all other radii spreading migth be important. If the shell is spreading then $\Delta$ in the above expressions should be replaced by $R/\gamma^2$. This delays the time at which the reverse shock reaches the shell and decreases the shell's density. These effects lead to a triple coincidence: $R_\Delta = R_\gamma = R_N$ with $\xi \approx 1$ and a mildly relativistic reverse shock during the period of effective energy extraction. Without spreading only a small fraction of the total energy is converted to thermal energy in the reverse shock. With spreading both shocks convert comparable amounts of energy.

2. $\xi < 1$ - the relativistic case: $R_N < R\gamma < R_\Delta < R_s$. The reverse shock becomes relativistic before it crosses the shell. Only a small fraction of the energy is converted at $R_\gamma$ and the kinetic energy is converted into internal energy only at $R_\Delta$. $R_s$ is larger than all other radii and spreading is unimportant. It is interesting to note that in this limit $\gamma_2(R_\Delta) \sim (l/\Delta)^{3/8}$ is independent of the initial Lorentz factor $\gamma$ and it is only weakly dependent on other parameters. This might have an important role in the fact that the observed radiation always appears as low energy $\gamma$-rays.

Neither the internal shocks nor the ISM inhomogeneity time scales are affected by these spherical considerations. The former depends only upon $\delta R$, $\Delta$ and $\gamma$ and the latter depends only on $l_{ISM}$ and $\gamma$. Both are constant throughout the spherical expansions.

## 4. Observational Implications to GRB

We estimate now the relevant parameters for GRBs and examine the possible relation between the observed time scales and the hydrodynamic time scales. We assume that the shocked material (either region 2 or 3) emits the radiation on a time scale shorter than the hydrodynamic time scales. A simple estimate of synchrotron cooling rate (assuming equipartition of the magnetic field energy) is consistent with this assumption.

The total energy of the bursts can be estimated directly from the observed fluxes (assuming cosmological distances) as: $E = 10^{51}$ergs. The ISM density has a typical value of: $n_1 = 1 \text{particle}/\text{cm}^3$. Only the ratio of these two quantities appears in our considerations and it determines the Sedov length $l \approx 10^{18}$cm.

The values of $\Delta$ and $\gamma$ are more ambiguous. It is known that $\gamma \geq 100$ in order for the shell to be transparent for $\gamma$-rays (Fenimore, Epstein & Ho, 1993; Woods & Loeb, 1995; Piran, 1995). A similar constraint can be obtained from the observed duration of the bursts and we adopt $\gamma = 10^3$ as our canonical value. The width of the shell is highly uncertain and relativistic effects allow it to be several order of magnitude larger than the common canonical value $\Delta = 10^7$cm. For these canonical parameters $\xi \cong 30 > 1$, corresponding to a Newtonian reverse shock. Nevertheless a value of $\xi \cong 0.1 < 1$ is also possible with reasonable parameters (for example $\Delta = 10^9$cm and $\gamma = 10^4$). Therefore both relativistic and Newtonian reverse shock are possible.

The bursts' duration is determined by the slowing down time of the shell. The emitting region moves towards the observer with a Lorentz factor $\gamma_2$. Two photons that are emitted with a time delay $dt$ will be detected with a time delay $dt/\gamma_2^2$. Additionally, an observer detects radiation from a region with an angular size $\gamma_2^{-1}$. A photon emerging from an angle $\gamma_2^{-1}$ away from the center of a region with a radius $R$ will be detected at a time $R/\gamma_2^2 c$ after a photon that emerges from the center (Katz, 1994). Thus, given a typical radius of energy conversion, $R_e$ the observed time scale is:

$$\Delta t_{obs} = R_e/\gamma_2^2 c = \begin{cases} \Delta/c & \text{if } \xi < 1 \quad \text{(Relativistic)}; \\ R_\gamma/\gamma^2 c \sim l/\gamma^{8/3} c & \text{if } \xi > 1 \quad \text{(Newtonian)}. \end{cases} \quad (11)$$

This time scale ranges from $\sim$ 1msec, for $\Delta = 3 \times 10^7$cm and $\gamma = 10^4$, to $\sim$ 100sec for $\gamma = 10^2$ and $\Delta = 10^{13}$cm (see fig. 1). In the Newtonian regime ($\xi > 1$) the observed time scale depends only on $\gamma$ while it depends only on $\Delta$ in the relativistic ($\xi < 1$) case. The observed durations of the brightest 30 bursts limit $\gamma$ to $100 < \gamma < 10^4$ with a typical value of $\approx 500$ and it limits $\Delta$ to

$\Delta < 3 \times 10^{12}$cm.

In the Newtonian case there appears a second time scale: the crossing time of the shell by the reverse shock. The corresponding observed time scale:

$$\tilde{t} \sim l^{3/4}\Delta^{1/4}/\gamma^2 c = t_{obs}/\sqrt{\xi} \;, \qquad (12)$$

is comparable to the measured time scale of variability in the bursts (10msec-10sec). However, it not clear if this time scale has any observational implications since only a small fraction of the energy is emitted by the reverse shock in this case (it might though produce primary photons that will be Compton scattered later by the forward shock). Furthermore, spreading prolongs this scale so that $\tilde{t} \approx t_{obs}$.

Another, more likely, source of the variability is inhomogeneity in the ISM. If the length scale of the inhomogeneity is $l_{ISM}$ and the density varies by one order of magnitude then the time scale for the observed variability will be:

$$t_{var} \sim l_{ISM}/10\gamma_2^2 c \qquad (13)$$

This time scale can be sufficiently short to produce the observed variability if $l_{ISM}$ is sufficiently small.

About 3% of the bursts contain precursors: weaker bursts that proceed the main burst. Two additional time scales appears here: the precursor's duration $\Delta t_{pre}$, and its separation from the main burst $\Delta t_{pre-maim}$. A natural explanation for the precursor phenomenon, within this model, is that it arises from internal shocks that take place at $R \approx \delta R\gamma^2 \leq \Delta\gamma^2$ while the main burst originates from the interaction with the ISM. (Mésźaros, & Rees; 1994) proposed that internal shock produce the main GRB while the interaction with the ISM produces the delayed GeV photons observed in some bursts). The duration of the precursor is $\Delta t_{pre} = \Delta/c$, which requires values of $\Delta$ as high as $10^{10} - 10^{12}$ cm to produce the observed precursors of $1 - 100$ sec. If $\xi > 1$ the main bursts will have a duration $\Delta t_{obs} \approx l/\gamma^{8/3}c = \xi^2 \Delta$. This will also be the typical separation $\Delta t_{pre-maim}$ between the precursor and the main burst. We expect a time delay between the precursor and the main burst which will be comparable to the duration of the main burst. Note that a correlation of the form $\Delta t_{pre-maim} \approx 4.5 \Delta t_{obs}$ exists (but was not reported) in the data of Koshut *et. al.*, (1995). If $\xi < 1$ then precursors do not occur (unless $\delta R \ll \Delta$) This is in agreement with the lack of observed precursors in short bursts.

## 5. Conclusions

We have calculated the hydrodynamic time scales of shocks during the interaction between an ultra-relativistic shell and the ISM. These time scales depend on the shock conditions which in turn depend only on energy and momentum conservations. Hence, we believe that these time scales are robust and independent of the unknown microphysics that takes place in these shocks. We find that with reasonable astrophysical parameters these time scales are in a good agreement with the observed time scales in GRBs. Our analysis shows that there are two kinds of shocks: Newtonian and Relativistic. The difference between them might correspond to the difference between the observed short and long bursts (Kouveliotou *et. al.* 1993). Finally, we suggest that precursors might be emitted due to internal shocks within the ultra-relativistic shell while the main burst emerges later from the interaction with the ISM.

We thank Ramesh Narayan and Peter Mészaros for valuable comments on this manuscript. This research was supported by BRF grant to the Hebrew University by NASA grant NAG5-1904 and by NSF grant PHY94-07194.

**Figure Captions**

Figure 1: The observed duration of the burst (dashed curve in the Newtonain regime and dotted curve in the relativistic regime) as function of $\gamma$ and $\Delta$ for $l = 10^{18}$cm ($E = 10^{51}$erg and $n_1 = 1 cm^{-3}$). The thick solid line ($\xi = 1$) separates

the Newtonian (lower left) and the Relativistic (upper right) regions. Spreading drives all Newtonian cases to the $\xi \approx 1$ line.

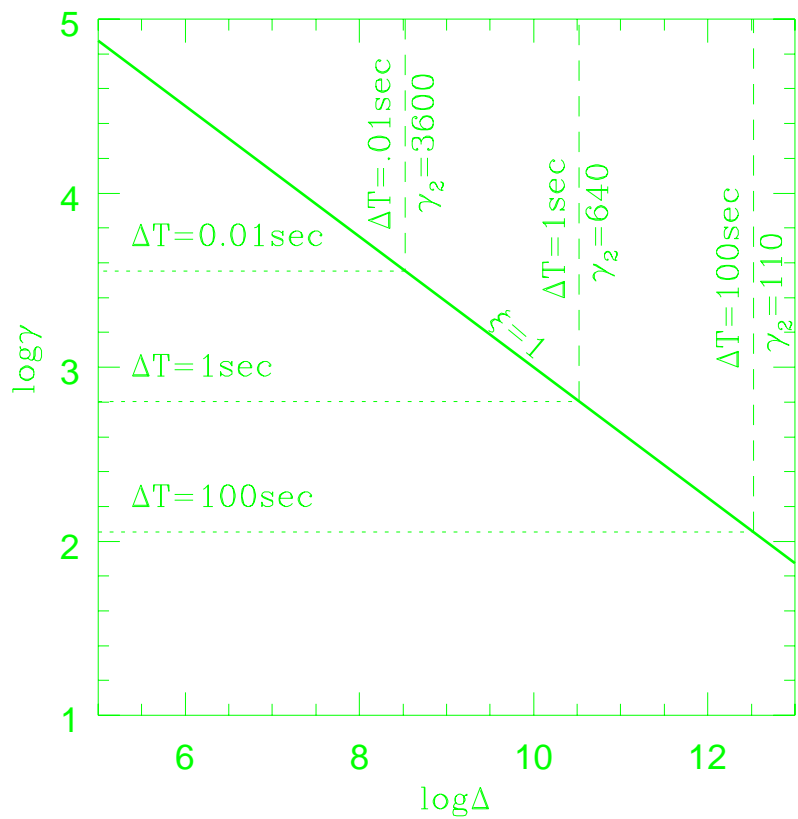